\documentclass[11pt]{article} 
\makeatletter
\def\draft{\textheight=10.5truein \textwidth=7.5truein \parindent=8pt
           \voffset=-1truein \topmargin=0Truein
           \ifcase \@ptsize \hoffset=-1.5truein \or \hoffset=-1.35truein
                        \or \hoffset=-1.15truein \fi}
\def\quality{\textheight=240mm \textwidth=160mm \topmargin=0Truein
             \ifcase \@ptsize \hoffset=-23mm
                     \or \hoffset=-20mm \or \hoffset=-15mm \fi}
\def\fps@figure{htbp} 
\makeatother%
\quality

\def\phi{\varphi}  \def\?#1{}     \def\n{\noindent} 
\def\bX{{\bf X}}   \def\bL{\bf L} \def\v{{\rm V}}  \def\n{\noindent}
\def\IZ{\hbox{{\rm Z}\kern-.3em{\rm Z}}}
\def\function#1{\left\{\!\!\!\begin{array}{ll} #1 \end{array} \right.}
\def\beq#1#2{\begin{equation} \label{#1} #2 \end{equation}}

\newcommand\filledsquare{\ \vrule width 1.5ex height 1.2ex} 
\def\qed{\hfill\filledsquare\linebreak\smallskip\par}

\def\mlbscale{1pt} 

\def\Bfig(#1,#2)#3#4{\begin{figure} \begin{center}
    \setlength{\unitlength}{\mlbscale} \begin{picture}(#1,#2) #3
    \end{picture} \end{center} \caption{#4} \end{figure}}

\def\bline(#1,#2)(#3,#4)(#5){\put(#1,#2){\line(#3,#4){#5}}}

\newtheorem{theorem}{Theorem}[section]       
\newtheorem{lemma}{Lemma}[section]           
\newtheorem{proposition}[lemma]{Proposition} 
\newtheorem{corollary}[lemma]{Corollary}     
\def\rmname{Remark}
\newtheorem{rmrk}[lemma]{\rmname}
\newenvironment{remark}{\begin{rmrk}\rm}{\end{rmrk}}     
             \catcode`@=11 \@addtoreset{equation}{section} \catcode`@=12
\def\proof{\smallskip \noindent {\bf Proof. \ }}

\begin{document}
\title{Variational principles in the analysis of traffic flows. \\
       (Why it is worth to go against the flow.)}
\author{Michael Blank\thanks{This research has been partially
                             supported by RFFI and INTAS grants.}
       \\ \\
        Russian Ac. of Sci., Inst. for
        Information Transmission Problems, \\
        B.Karetnij 19, 101447, Moscow, Russia, blank@iitp.ru}
\date{17.01.2000}
\maketitle

\begin{abstract} By means of a novel variational approach and using
dual maps techniques and general ideas of dynamical system theory
we derive exact results about several models of transport flows,
for which we also obtain a complete description of their limit
(in time) behavior in the space of configurations. Using these
results we study the motion of a speedy passive particle (tracer)
moving along/against the flow of slow particles and demonstrate
that the latter case might be more efficient.
\end{abstract}

\bigskip%
\n{\bf Keywords}: traffic flow, dynamical system, variational
principle, skew product map.

\bigskip%
\n{\bf AMS Subject Classification}: Primary 37E15; Secondary
37D35, 37B15, 60K.

\section{Introduction}\label{section-intro}
During the last decade problems related to transport in complex
systems attracted a huge amount of interest in particular due to
their evident practical importance. In this paper we deal with
theoretical aspects of phenomena arising in the modeling of
highway traffic flow. Previously much of the effort in the
construction and analysis of such models was concentrated on
discrete (on time and on space) stochastic models introduced in
\cite{NS} and later studied by many authors (see \cite{BF} for
review and further references). All these models were based on
the idea to describe the dynamics in terms of cellular automata
and to a large extent were studied by means of numerical
simulation (especially because of low computational cost of the
numerical realization of cellular automata rules, which made it
possible to realize large-scale real-time simulations of urban
traffic \cite{SN}).

My own interest to this type of problems is mainly due to the
following practical observation. Going by foot in a slowly moving
crowd it is faster to go against the ``flow'' than in the same
direction as other people go. This effect is especially
pronounced if one is moving near a boundary between two ``flows''
of people going in opposite directions. A standard probabilistic
model of a diffusion of a particle against/along the flow clearly
contradicts to this observation, which very likely indicates a
very special (nonrandom) intrinsic structure of the flow in this
case. The main aim of the present paper is to study how this
structure emerges from arbitrary (random) initial configurations
in some simple models of the traffic flow.

The main quantity of interest in traffic models is the average
velocity of cars $\v$ and its dependence on the density of cars
$\rho$ (called a fundamental diagram) is typically studied in the
steady state. Various approaches starting from the mean-field
approximation \cite{KS} to combinatorial techniques and
statistical mechanics methods \cite{F} were used in the analysis
of this type of models. In what follows we shall restrict the
consideration to deterministic discrete time and space traffic
models. The simplest model (among that we consider) describes the
dynamics of cars moving along a one-row motor road and is defined
as follows.

The road is associated to a one-dimensional ordered lattice of
size $N$ with periodic boundary conditions and each position on
the lattice is either occupied by a particle, or empty. Denote
the number of particles in the configuration by $m$. Then the
density of particles $\rho$ is equal to $\frac{m}{N}$. On the
next time step each particle remains on its place if the next
position is occupied, and moves forward by one place otherwise.
We shall call this model the traffic model with slow particles to
distinguish it from other ones.

Quite recently in \cite{BF,F} this model (and some its
generalizations) was studied by means of cellular automata
techniques. One of the most intriguing phenomenon related to this
model is a drastic change of the shape of the curve describing
the dependence $\v(\rho)$ of the average velocity of particles on
their density in the steady state when the density passes the
$1/2$ value (see Fig.~\ref{av-speed-den}). It was found
numerically and confirmed analytically in the limit of large $N$
(and for a typical initial configuration) \cite{BF,F} that
$\v(\rho)$ is equal to $1$ while $\rho<1/2$ and then goes down to
zero as $\frac1\rho-1$ for $\rho\ge1/2$.
\Bfig(150,100)
      {\footnotesize{
       \bline(0,0)(1,0)(150)   \bline(0,0)(0,1)(100)
       \bline(0,100)(1,0)(150) \bline(150,0)(0,1)(100)
       \bline(20,20)(1,0)(120) \bline(20,20)(0,1)(70)
       \bline(20,55)(1,0)(55)  \bezier{50}(75,55)(90,23)(130,20)
       \put(135,23){$\rho$}    \put(73,10){$\frac12$}  \put(125,12){1}
       \put(12,85){$\v$}       \put(10,52){$1$}
       \put(95,37){$\frac1\rho-1$}
      }}
{Dependence of the average velocity of particles on their density.
\label{av-speed-den}}

Despite an apparent simplicity of the model with slow particles
its dynamics (especially during the transient period and/or in
the high density case) is rather nontrivial. The description of
the dynamics in terms of cellular automata makes it possible
(using a rather complicated combinatorial techniques) to derive
an asymptotic description in the limit of large $N$ \cite{F}. 
Another approach based on the ultradiscrete limit of the Burgers 
equation was proposed in \cite{NT}, where the dependence $\v(\rho)$ 
was proven for a lattice of size $N$ but with the estimate of the 
transient period $N/2$, which rules out the generalization for 
the case of the infinite lattice and nonperiodic initial 
configurations. In this paper we consider this model from a bit 
more general point of view as a discrete time dynamical system 
(map $T$) acting in the space of all possible configurations 
$\bX$ -- collections of zeros and ones (describing the positions 
of particles). We derive the variational approach based on the 
observation that the average velocity of any configuration does 
not decrease in time (Proposition~\ref{var-pr}). Simultaneously 
to the dynamics of particles one can study the dynamics of empty 
places. Observe that each of these dynamics determines the other 
one in the unique way. To make use of this observation we introduce 
a dual map $T^{*}$ corresponding to the dynamics of empty places. 
By means of these two basic ideas (variational approach and dual
maps techniques) we first prove the formula for the dependence of
the average velocity on the density for any (may be small) finite
lattice lengths $N$ and any (not necessary typical) initial
configurations. We find also that steady state configurations
demonstrate certain periodic in time patterns, whose features are
described in Theorem~\ref{1D-th} as well as the convergence to
the steady state and the duration of the transient period.
Qualitatively our main result about this model is that the
following alternative takes place: either the flow consists of
only free particles (i.e. there are no clusters of particles), or
there are no clusters of empty places.

The paper is organized as follows. In Section~\ref{section-1D-per} we
study in detail the above formulated simplest model, which we call 1D
periodic model with slow particles. In Section~\ref{sect-1D-unb} we
consider the same model but on the unbounded lattice, which
significantly changes the dynamics and cannot be obtained in the
limit of large $N$ from the previous one. To demonstrate the power of
our dual maps techniques we introduce in Section~\ref{sect-1D-speedy}
the model of the one-row motor road with speedy cars. The latter
means that instead of the moving by at most 1 position, a particle
moves forward until the next occupied position. We study both periodic
and unbounded 1D cases. From the mathematical point of view the main
difference of this model from the previous one is that the dynamics
is not local, which makes it impossible to describe this model in
terms of cellular automata. In Section~\ref{sect-2D} we generalize
the traffic models with slow particles for a more practically
interesting case of a multi-row motor road and study its properties.
Finally in Section~\ref{sect-tracer} we discuss a model of a passive
tracer in the flow generated by the traffic model with slow particles
confirming our practical observation above.

It is worth mention that to the best of our knowledge only the
simplest 1D periodic traffic model with slow particles (among the
models considered in the paper) was discussed in the literature
previously. For some missing definitions related to dynamical
systems theory (especially for systems acting on discrete phase
spaces) we refer the reader to books on ergodic theory of
dynamical systems (see, for example, \cite{Bl20}).

\section{1D traffic model with slow particles on the finite lattice}
\label{section-1D-per} Let $\bX=\{0,1\}^{N}$ be the set of all
possible configurations -- collections $X$ of $N$ elements from
the alphabet of 2 letters $0$ and $1$. We consider a map
$T:\bX\to\bX$ defined as follows:%
\beq{1D-map}{TX(x)
  := \function{ 1 &\mbox{if } X(x)=0 \mbox{ and } X(x-1)=1\\
                1 &\mbox{if } X(x)=1 \mbox{ and } X(x+1)=1\\
                0 &\mbox{otherwise}.} }
We assume here periodic boundary conditions, i.e. $N+1\equiv1$
and $0\equiv N$. Observe that this map is not one-to-one and thus
the backward (in time) dynamics cannot be defined in a unique
way. See examples of the dynamics under the action of the map $T$
on Fig.~\ref{1D-ex-fig}.

\begin{figure} \begin{center}
(a)            \qquad\qquad \qquad  (b)            \par
0011011100010  \qquad  t=0  \qquad  1011011100110  \par
0010111010001  \qquad  t=1  \qquad  0110111010101  \par
1001110101000  \qquad  t=2  \qquad  1101110101010  \par
0101101010100  \qquad  t=3  \qquad  1011101010101  \par
0011010101010  \qquad  t=4  \qquad  0111010101011  \par
0010101010101  \qquad  t=5  \qquad  1110101010110  \par
1001010101010  \qquad  t=6  \qquad  1101010101101  \par
0100101010101  \qquad  t=7  \qquad  1010101011011  \par
\end{center} \caption{Two examples of the dynamics of the model with
slow particles: (a) $m=6, \; N=13, \; \rho=6/13<1/2$, \ \
(b) $m=8, \; N=13, \; \rho=8/13>1/2$. \label{1D-ex-fig}}
\end{figure}

We shall say that there is a particle at a position $x$ on the
lattice if $X(x)=1$ and that this position is empty otherwise. A
particle at a position $x$ is called free if $X(x+1)=0$. A group
(more than 1) of consecutive particles (empty places) we call a
cluster of particles (empty places). Observe that the number of
particles is preserved under dynamics. Under the action of the
map $T$ on the next time step each particle will either go
forward by 1 place if this place is empty (occupied by 0) or will
remain on the same place otherwise. Introducing the notion of a
local velocity:
$$ v(X,x) := \function{ 1 &\mbox{if } X(x)=1 \mbox{ and } X(x+1)=0\\
                        0 &\mbox{otherwise} ,}$$
we define the average (in space) velocity of (particles in) the
configuration $X$ as
$$ \v(X) := \frac1{m(X)} \sum_{x} v(X,x) ,$$
where $m(X)\equiv\sum_{x}X(x)$ is the total number of particles
in the configuration. Observe that $\v(X)$ is equal to the number
of free particles divided by the total number of particles in the
configuration.

\begin{theorem} \label{1D-th} For any $N$ and any initial
configuration with $m \le N$ particles after at most
$\min(m,N-m)$ iterations the configuration will become periodic
(in time) with the period $N$ and the average velocity
$\v=\min(1,\frac{N}{m}-1)$.
\end{theorem}

\begin{remark} The smallest period (in time) of the periodic
configuration above might be smaller than $N$, indeed, for $N=2m$
the smallest period is equal to $2$, i.e. there exists $X\in\bX$
such that $T^{2}X=X$.
\end{remark}

The proof of the theorem consists of the following lemmas.

\begin{lemma} \label{cluster-length}
The length of any cluster of particles cannot increase, and the
number of free particles cannot decrease.
\end{lemma}

\proof Fix a cluster of particles. On the next time step its
first particle (with the largest $x$-coordinate) goes out of the
cluster and at most one particle can join the cluster from
behind. If the number of particles is 2 and no particle will join
the cluster from behind on the next time step, only one particle
will remain in it, and according to our definition the cluster
disappears.

Consider now the number of free particles. On the next time step
each cluster of particles loses the first its particle which
becomes a free one and at most the same number of free particles
can join clusters coming from behind. Therefore the number of
free particles cannot decrease. \qed

\begin{corollary} Traffic jams cannot appear from nothing.
\end{corollary}

On the other hand, since the average velocity of a configuration
is equal to the number of free particles in it and by
Lemma~\ref{cluster-length} this number grows in time we come the
the following variational principle.

\begin{proposition} (Variational principle) \label{var-pr}%
The functional $\v(X)$ (average velocity) increases under the
dynamics up to the moment when it takes its maximal possible
value. \end{proposition}

\begin{lemma} Any configuration after a finite number of time steps
gets into a periodic one. \end{lemma}

\proof The phase space $\bX$ of the considered dynamical system
$(\bX, T)$ is finite and therefore any its trajectory $T^{t}X$
will begin repeating after a finite number of iterations. Thus
any limit set of the map $T$ consists of periodic configurations.
\qed

The map $T$ describes the dynamics of particles. It turns out
that often it is simpler to study the dynamics of empty places
instead. To make use of this observation for a configuration $X$
we introduce a dual one $X^{*}(x):=1-X(x)$ for all $x$ and define
a dual map $T^{*}$ acting on the same space $\bX$:
\beq{1D-map-dual}{T^{*}X(x)
  := \function{ 0 &\mbox{if } X(x)=1 \mbox{ and } X(x+1)=0\\
                0 &\mbox{if } X(x)=0 \mbox{ and } X(x-1)=1\\
                1 &\mbox{otherwise},} }
assuming again periodic boundary conditions. One can easily see
that the map $T^{*}$ describes the motion of empty places under
the action of the map $T$, which in this case satisfies the same
rules as the the motion of particles except it goes in the
backward direction in space. A direct computation gives the
following representation, which can be considered as a definition
of the dual map in the general case (not only for this specific
model).

\begin{lemma}\label{dual-def} $TX = (T^{*}X^{*})^{*}$. \end{lemma}

\begin{corollary}\label{cor-dual}
All results for the map $T$ hold true also for the dual map and
vice versa.
\end{corollary}

\n {\bf Proof} of Theorem~\ref{1D-th}. Assume first that the
density of particles $\rho(X):=\frac{m}{N}$ in the initial
configuration $X$ is less or equal to $1/2$. If all the particles
in this configuration are free then each particle moves with the
velocity $1$ and the trajectory starting from this configuration is
periodic (in time) with (may be not minimal) period $N$. Indeed,
after $N$ iterations each particle will return to its initial
position.

If there are clusters of particles in $X$ we need to show that
after at most $m$ iterations all particles will become free.
Indeed, by Lemma~\ref{cluster-length} the number of free particles 
does not decrease. Denote by $\tilde m$ the length of the largest 
cluster of empty places. Let us prove by induction on $m$ that 
for any pair of positive integers $m,n$ such that $m\le n/2$ 
and for any initial configuration after at most 
$t_c := \min(m, \tilde m)$ iterations all particles will become 
free. This statement is trivial for $m=1$ and $n\ge2$. Assuming 
that it holds for some $m$ let us prove it for $m+1$. Let $X$ be a 
configuration with $m+1$ particle on the lattice of size $n\ge2(m+1)$
and let $x_0$ be the position of the first particle after (one of) 
the largest clusters of empty spaces of length $\tilde m$. Fix 
this particle and consider the dynamics of others. By the 
induction hypothesis during the first $\min(m, \tilde m)$ 
iterations other particles do not collide with the chosen one. 
Therefore their dynamics is the same as if there are only $m$ 
particles. Again by the induction hypothesis all these particles 
will become free after $\min(m, \tilde m)$ iterations and at that 
moment either the chosen particle is free also, or it forms a 
cluster of size 2. Therefore after the next iterations it becomes 
free as well. 

It remains to analyze initial configurations with the density of
particles greater than $1/2$. This situation is much more
complex, because any configuration of this type contains
clusters, which are interchanging particles between themselves
and never disappear completely. To overcome this difficulty we
consider a dual map $T^{*}$. The density of empty places is equal
to $\frac{N-m}N<1/2$. Therefore by Corollary~\ref{cor-dual} and
the first part of the proof under the action of the map $T^{*}$ a
dual configuration $X^{*}$ after at most $t_{0}=N-m$ iterations
will get into a periodic configuration $(T^{*})^{t_{0}}X^{*}$
consisting of $N-m$ free particles with the period (in time) $N$
for the map $T^{*}$. Clearly for the dual to it configuration we
have the following identity $((T^{*})^{t_{0}}X^{*})^{*} =
T^{t_{0}}X$ and the latter configuration is $T$-periodic with the
same period. To finish the proof we calculate the average velocity:
$$ \v(T^{t_{0}}X) = \frac{N-m}m = \frac{N}m-1 .$$
\qed

It is worth notice that our estimate of the duration of the
transient period $\min(m,N-m)$ is exact: if $m<N-m$ and the
initial configuration consists of only one cluster of length $m$
the duration of the transient period is equal to $m$.

\begin{corollary} Qualitatively our main result about this model
is that the following alternative takes place: either the flow
consists of only free particles (i.e. there are no clusters of
particles), or there are no clusters of empty places.
\end{corollary}

On the other hand, there is no a priori information about the
distribution of lengths of clusters of particles because this
distribution depends on the initial configuration and can differ
between various periodic limiting configurations.

\section{1D traffic model with slow particles on the unbounded lattice}
\label{sect-1D-unb} Consider now the unbounded one-dimensional
case, i.e. $\bX:=\{0,1\}^{\IZ^1}$ and the map $T$ is defined by
the formula (\ref{1D-map}) in the same way as in the previous
section, except for the periodic boundary conditions. For a
configuration $X\in\bX$ we define the notion of a
subconfiguration $X_{k}^{n}:=\{X(k),X(k+1),\dots,X(n)\}$, i.e a
collection of entries of $X$ between the pair of given indices
$k<n$, and introduce the corresponding density and the average
velocity:
$$ \rho(X_{k}^{n}) := \frac{m(X_{k}^{n})}{n-k+1}, \qquad
     \v(X_{k}^{n}) := \frac1{m(X_{k}^{n-1})} \sum_{x=k}^{n-1}v(X,x) ,$$
where $m(X_{k}^{n})$ stays for the number of particles in the
subconfiguration $X_{k}^{n}$.

By the density and the average velocity (of particles) of a entire
configuration $X\in\bX$ we mean the following limits (if they are
well defined):
$$ \rho(X) := \lim_{n\to\infty}\rho(X_{-n}^{n}), \qquad
     \v(X) := \lim_{n\to\infty}\v(X_{-n}^{n}) ,$$
otherwise one can consider the corresponding upper and lower
limits, which we denote by $\rho_{\pm}(X)$ and $\v_{\pm}(X)$.

Notice that in distinction to the finite case these quantities
make sense not for all possible configurations. We shall say that
a configuration $X$ satisfies the regularity assumption (or
simply regular) if there exists a number $\rho$ and a monotonous
one-to-one function $\phi(n) \to 0$ as $n\to\infty$, such that
for any $n\in\IZ^{1}$, $N\in\IZ_{+}^{1}$ and any subconfiguration
$X_{n+1}^{n+N}$ of length $N$ the number of particles in this
subconfiguration $m(X_{n+1}^{n+N})$ satisfies the inequality %
\beq{reg-ineq}{
 \left| \frac{m(X_{n+1}^{n+N})}N - \rho \right| \le \phi(N) .}%
It is clear that at least for a configuration $X$ satisfying the
regularity assumption the density $\rho(X)$ is well defined and
is equal to the value $\rho$ in the formulation of the
assumption.

\begin{theorem}\label{1D-unb-th} Let the initial configuration
$X$ satisfies the regularity assumption with $\rho\ne1/2$. Then
after a finite number of iterations the average velocity of
particles becomes equal to $\min(1, \frac1\rho-1)$.
\end{theorem}

\proof Notice that Lemma~\ref{cluster-length} holds true in this
case also. Moreover it can be applied for the dual map as well
and thus the length of any cluster of empty places cannot
decrease. This shows that the variational principle
(Proposition~\ref{var-pr}) holds in this case also. Moreover it
can be reformulated to take care about configurations for which
neither the density, nor the average velocity are well defined.

\begin{proposition} (Variational principle) \label{var-pr-unb}
For any configuration $X$ any pair of its particle denote by
$x'(t) < x''(t)$ their positions at the moment $t$. Then the
average velocity $\v(X_{x'(t)}^{x''(t)})$ of the subconfiguration
$X_{x'(t)}^{x''(t)}$ increases monotonically with $t$ up to the
moment (may be infinite) when it takes its maximal possible
value. \end{proposition}

The basic technical step of the proof of Theorem~\ref{1D-unb-th}
is given by the following statement.

\begin{lemma}\label{ref-inv} If $X$ satisfies the regularity
assumption, then the same holds true for $T^{t}X$ for any $t>0$.
\end{lemma}

\proof Clearly it is enough to prove this statement for $t=1$.
Assume that this is not true and for some $N > \phi^{-1}(\rho)$
there is a subconfiguration $(TX)_{n+1}^{n+N}$ of length $N$ in
the configuration $TX$ such that
$$ m((TX)_{n+1}^{n+N}) < (\rho - \phi(N))N .$$
This can happen only if the following equalities are satisfied
simultaneously
$$ m(X_{n+1}^{n+N})=(\rho - \phi(N))N, \quad X(n)+X(n+N+1)=0,
   \quad X(n+N)=1, $$
i.e. on the next time step no particle from behind will come to
this interval and there is a free particle in the last position
of the considered interval, which leaves it. On the other hand, from
these equalities we immediately deduce that
$$ m(X_{n}^{n+N-1}) = (\rho - \phi(N))N - 1 ,$$
which contradicts the regularity assumption. In the same way one
can prove that the inequality
$$ m(X_{n}^{n+N-1}) \le (\rho + \phi(N))N $$
cannot break as well. \qed

Now we can return to the proof of Theorem~\ref{1D-unb-th}. Assume
first that $\rho<1/2$. Denoting by $[\cdot]$ the integer part of
a number, we get that for any $N \ge
N_{c}:=[\phi^{-1}(\frac12-\rho)]$ the number of particles in any
subconfiguration $(T^{t}X)_{n+1}^{n+N}$ can be estimated as
$$ m((T^{t}X)_{n+1}^{n+N})
  \le (\rho+\phi(N))N \le (\rho+\phi(N_{c}))N \le N/2 .$$
Applying the same machinery as in the previous section one can
show that after at most
$$ t_{c} := (\rho+\phi(N_{c}))N_{c} \le \frac{N_{c}}2
   \le \frac12 \phi^{-1}(\frac12-\rho) $$
iterations all particles will become free ones, which implies the
unit average velocity.

In the remaining case $\rho>1/2$ we follow the same idea as in
the proof of Theorem~\ref{1D-th} and pass to the dual map and the
dual configuration. Observe that if a configuration $X$ satisfies
the regular assumption with the density $\rho$ and the rate
function $\phi$, then the dual configuration $X^{*}$ also
satisfies it with the density $1-\rho$ and the same rate function
$\phi$. Indeed $m((X^{*})_{n+1}^{n+N}) = N - m(X_{n+1}^{n+N})$
for any $n,N$ and thus
$$ \left| \frac{m((X^{*})_{n+1}^{n+N})}N - (1-\rho) \right|
 = \left| \frac{m(X_{n+1}^{n+N})}N - \rho) \right| \le \phi(N) .$$
On the other hand, the density of the dual configuration
$1-\rho\le1/2$ which, having in mind that the only difference
between the maps $T$ and $T^{*}$ is the direction of motion,
gives us the possibility to apply the first part of the proof.
\qed

Observe that in the proof of Theorem~\ref{1D-unb-th} we actually
derived an estimate of the length of the transient period as
$t_{c}:=(\rho+\phi(N_{c}))N_{c}$ which goes to infinity as
$\rho\to1/2$. This is the reason why Theorem~\ref{1D-unb-th} does
not cover the boundary case $\rho=1/2$, which we discuss below.

\begin{theorem}\label{1D-unb-half-th} Let the initial configuration
$X$ satisfies the regularity assumption with $\rho=1/2$ and let
$x'(t) < x''(t)$ be positions of two fixed particles at the
arbitrary moment $t$. Then the average velocity of the
subconfiguration $X_{x'(t)}^{x''(t)}$ converges to $1$ as
$t\to\infty$.
\end{theorem}

\proof Choose an integer $\hat N$ and consider a configuration
$\hat X$ obtained from the configuration $X$ by the following
operation: for each $k$ we remove from the configuration $X$ the
closest from behind particle to the position $k \hat N$.
$$   \left|\frac{m(\hat X_{n+1}^{n+N})}N - (\rho-\frac1{\hat N}) \right|
 \le \left|\frac{m(X_{n+1}^{n+N})}N - \rho \right|
   + \left|\frac{m(\hat X_{n+1}^{n+N})}N
         - \frac{m(X_{n+1}^{n+N})}N + \frac1{\hat N} \right| 
 \le \phi(N) + \frac1{\hat N} .$$
Thus the configuration $\hat X$ is also regular but with the
density $\frac12-\frac1{\hat N} < \frac12$. Thus according to
Theorem~\ref{1D-unb-th} after a finite number of iterations
$t_{c}$ the average velocity of the configuration $T^{t_{c}}\hat
X$ becomes equal to $1$.

Making an opposite operation, namely inserting a particle to the
configuration $X$ to the closest from behind to $k\hat N$ empty
position for each $k$, we obtain a regular configuration with the
density $\frac12+\frac1{\hat N} > 1/2$. Again by
Theorem~\ref{1D-unb-th} after a finite number of iterations the
average velocity of this configuration becomes equal to
$$ \frac1{\frac12-\frac1{\hat N}}-1 = 1 + \frac4{\hat N - 2} \to 1
   \quad {\rm as} \quad \hat N \to\infty .$$

Thus both (arbitrary close as $\hat N\to\infty$) approximations
to the configuration $X$ have after a finite number of iterations
(depending on $\hat N$) the average velocity deviating from $1$ by
$O(\hat N)$. It remains to show that the average velocity of a
subconfiguration of the configuration $X$ can be estimated from
above and from below by those from above approximations. Let $X$
and $Y$ are two configurations such that $X(x)\le Y(x)$ for all
$x$ and let $x'(t) < x''(t)$ be positions of two fixed particles
in the configuration $X$ at the arbitrary moment $t$. Denote by
$y'(t) < y''(t)$ positions of the same particles in the
configuration $Y$. Then
$$ \v(X_{x'(t)}^{x''(t)}) \ge \v(X_{x'(t)}^{x''(t)}) $$
for any moment of time $t$. Indeed, additional particles in the
configuration $Y$ present only obstacles to the motion of other
particles, thus making the average velocity slower (or at least
not faster). \qed

\section{1D traffic model with speedy particles}\label{sect-1D-speedy}
The dual map $T^{*}$ that we made use of in the previous sections
was mirror symmetric with respect to the map $T$. In the model
considered in this section we show that the dual map might have a
different structure as well.

Let us start with the one-dimensional finite periodic case, but
with speedy particles instead of particles moving with the velocity
$1$ as above. This means that if in front of a particle there are
exactly $n$ empty places it moves by $n$ places forward and
remains in the same place if the next place is occupied. By the
average velocity of (particles in) the configuration $X$ we mean
the total distance covered by the particles from $X$ during the
next time step divided by the number of particles. The
corresponding map of the space $\bX=\{0,1\}^{N}$ into itself we
again denote by $T$. See examples of the dynamics under the
action of the map $T$ on Fig.~\ref{1D-ex-fig-speedy}.
\begin{figure} \begin{center}
(a)            \qquad\qquad \qquad  (b)            \par
0011011100010  \qquad  t=0  \qquad  1011011100110  \par
0110111000100  \qquad  t=1  \qquad  0110111001101  \par
1101110001000  \qquad  t=2  \qquad  1101110011010  \par
1011100010001  \qquad  t=3  \qquad  1011100110101  \par
0111000100011  \qquad  t=4  \qquad  0111001101011  \par
1110001000110  \qquad  t=5  \qquad  1110011010110  \par
1100010001101  \qquad  t=6  \qquad  1100110101101  \par
1000100011011  \qquad  t=7  \qquad  1001101011011  \par
\end{center} \caption{Two examples of the dynamics of the model with
speedy particles: (a) $m=6, \; N=13, \; \rho=6/13<1/2$, \ \
(b) $m=8, \; N=13, \; \rho=8/13>1/2$. \label{1D-ex-fig-speedy}}
\end{figure}

\begin{theorem}\label{1D-speedy-th} The dual map in this case
satisfies the relation $T^{*}X^{*}=TX^{*}$, and for any
nontrivial (having both zeros and ones) initial configuration $X
\in \bX$ the average velocity does not depend on time and is equal
to $\frac{N}{m(X)}-1$, where $m(X)$ is the number of particles in
the configuration $X$.
\end{theorem}

\proof We consider this result as an illustration of the dual
maps techniques and and a base for the introduction of the
passive tracer model considered in Section~\ref{sect-tracer}.
Therefore we do not give explicit representations for the maps
$T, T^{*}$ and present only a sketch of the proof in this case.

The relation between the map $T$ and the dual one $T^{*}$ can be
checked by a direct computation (which we leave for the reader).
Notice that unlike the dual map in Section~\ref{section-1D-per}
the dynamics of empty places is exactly the same as the dynamics
of particles and occurs in the same direction in space.

According to the definition each particle on the next time step
moves forward by the number of empty places in front of it.
Therefore the total number of places to which the particles in
the configuration $X$ move forward is equal to the number of
empty places. Thus the average velocity is equal to $\v(X) =
\frac{N-m(X)}{m(X)} = \frac{N}{m(X)}-1$.

Using the same argument as in the proof of Theorem~\ref{1D-th} we
deduce that any trajectory of the map $T$ gets periodic after a
finite number of iterations and a trivial observation that the
number of empty places is invariant under dynamics finishes the
proof. \qed

Consider now the model of speedy particles on the unbounded 1D
lattice $\IZ^{1}$ in analogy to the model in
Section~\ref{sect-1D-unb}. The map $T$ is defined as above and
acts in the space $\bX=\{0,1\}^{\IZ^{1}}$. We shall say that a
configuration $X \in \bX$ satisfies the Law of Large Numbers if
for any integer $n$ the limit
$$ \lim_{N\to\infty} \frac{m(X_{n+1}^{n+N})}N $$
is well defined and does not depend on $n$. Clearly the above
limit (if it exists) coincides with the density of particles
$\rho(X)$ in the configuration $X$.

The dual map in the considered case satisfies the same relation
$T^{*}X^{*}=TX^{*}$ as in Theorem~\ref{1D-speedy-th} and the
dynamics is described by the following statement.

\begin{theorem}\label{1D-speedy-th-unb} Let an initial configuration
$X \in \bX$ satisfy the Law of Large Numbers and let
$\rho(X)\cdot(1-\rho(X))\ne0$. Then the average velocity does not
depend on time and is equal to $\frac1{\rho(X)}-1$.
\end{theorem}

\proof Let $x'<x''$ be positions of two particles in the
configuration $X$. Then the average velocity of the subconfiguration
$X_{x'}^{x''}$ is equal to the number of empty places in this
subconfiguration, divided by the number of particles in it, i.e.
$$ \v(X_{x'}^{x''})
 = \frac{x'' - x' - m(X_{x'}^{x''})}{m(X_{x'}^{x''})}
 = \frac{x'' - x'}{m(X_{x'}^{x''})} .$$
Now taking into account that this relation holds for arbitrary
positions of particles, we immediately get the desired statement.
\qed

\begin{remark} It is of interest that the average velocity in both
discussed speedy particles models coincides with the average
velocity in the models with slow particles in the case of high
density (i.e. when $\rho(X)>1/2$). The explanation is that in the
models with slow particles with the density $\rho(X)>1/2$ the
typical distance between particles (in the steady state) is $0$
or $1$, which makes no difference between the dynamics of speedy
and slow particles.
\end{remark}

Observe that in the models with speedy particles the dynamics is
richer than in the models with slow particles, for example the
statement of Lemma~\ref{cluster-length} does not hold, i.e.
clusters of particles can grow and traffic jams become typical
even for the case of low density of particles.

\section{2D traffic model}\label{sect-2D}
 From the point of view of real traffic problems a clear shortage
of all models considered above is the absence of the possibility
to go around particles staying in a traffic jam. Indeed in a
one-row motor road (which was the main considered example) this
is not possible. To overcome this restriction we consider a model
describing the motion of slow particles on a multi-row motor
road.

A lattice in our model is a $N\times K$-strip, describing a
one-way cyclic road of length $N$ consisting of $K$ parallel
rows. We denote the first coordinate corresponding to the spread
of the road by $x \in \{1,\dots,N\}$, and the second one (the row
number) by $y \in \{1,\dots,K\}$, and assume periodic boundary
conditions on $x$. In terms of particles the dynamics is defined
as follows. If there is a particle at a position $(x,y)$ (i.e.
$X(x,y)=1$) then this particle moves forward by one place to the
position $(x+1,y)$ if this place is not occupied, else moves to
the left to the position $(x,y+1)$ if this place and the place
before it are empty, else moves to the right to the position
$(x,y-1)$ if this place and the place before and the next place
to the right are not occupied, and remains on its place
otherwise. To be consistent we assume that the $0$-th, $(K+1)$-th
and $(K+2)$-th (virtual) rows are completely occupied. Recall
that the space is $N$-periodic on the $x$-coordinate.

Observe that this model is nothing more than the simplest
formulation of the standard traffic rules.

The space of all possible configurations of this model is
$\bX=\{0,1\}^{N K}$ and the corresponding map describing the
dynamics of the configurations in this space can be written as
follows: \beq{2D-map}{TX(x,y) := \function{
      1 &\mbox{if } X(x,y)=0 \mbox{ and } X(x-1,y)=1\\
      1 &\mbox{if } X(x,y)=1 \mbox{ and } X(x+1,y)=1\\
      1 &\mbox{if } X(x,y)+X(x-1,y)=0 \\
        &\quad      \mbox{ and } X(x,y+1)\cdot X(x+1,y+1)=1\\
      1 &\mbox{if } X(x,y)+X(x-1,y)+X(x,y+1)=0 \\
        &\quad      \mbox{ and } X(x,y-1)\cdot X(x+1,y+1)=1\\
      0 &\mbox{otherwise}.} }
The dual map for this model can be easily defined and is
described by the following statement. Again to be consistent we
need to add an additional $(-1)$-th completely occupied virtual
row to the lattice.

\begin{lemma} The dual map $T^{*}$ satisfies the general
statement of Lemma~\ref{dual-def} and the explicite formula for
$T^{*}X(x,y)$ can be obtained from the relation (\ref{2D-map}) by
changing everywhere $x+1$ by $x-1$, $x-1$ by $x+1$, and $y+1$ by
$y-1$.
\end{lemma}

This statement can be checked by a direct computation.

\begin{corollary} The dynamics of empty places under this model
is the same as the dynamics of particles except it occurs in the
opposite direction in space along the $x$-coordinate, i.e. the
correspondence between the dynamics is the mirror symmetry.
\end{corollary}

We are interested in the average velocity along the
$x$-coordinate, which we define in the same way as in the
previous sections. From the first sight it seems that the
presence of additional possibilities to go to the right or to the
left (if the way forward is blocked) should improve the traffic.
The following statement shows that this is not the case and
moreover in the 2D case the average velocity might be much slower
than the velocity predicted by the 1D model with slow particles.

\begin{theorem}\label{2D-th} For any initial configuration $X$
in the case of the periodic bounded lattice and any regular
initial configuration $X$ with $\rho\ne1/2$ in the case of the
unbounded (on the $x$-coordinate) lattice after a sufficiently
large number of iterations the upper limit for the average
velocity satisfies the inequality
$$ \v_{+}(T^{t}X) \le \min(1, \frac1{\rho(X)}-1), $$
while the lower limit can be arbitrary close to $0$ even in the
case of low density of particles ($\rho<1/2$).
\end{theorem}

Notice that in the bounded case exactly as in the 1D model on the
bounded lattice any initial configuration becomes periodic after
a finite number of iterations. Moreover in this case both limits
$\v_{\pm}$ coincide.

\proof We start with the case $\rho\le1/2$ and the periodic (on
the $x$-coordinate) bounded lattice of size $N K$. Our first aim
is to show that for any number $\rho\in[0,1/KN,\dots,1]$ a
configuration of the density $\rho$ cannot have the average velocity
larger than the one predicted by the $1D$ model with slow
particles. If $\rho N \le [N/2]$, then one can construct a
configuration $X_{+}$ consisting of exactly $\rho N$ free
particles on each of $K$ rows. Clearly the configuration $X_{+}$
is $N$-periodic and has the maximal possible average velocity
$1$. Otherwise at least in one of the rows there is a cluster of
particles and hence the average velocity is strictly smaller than
$1$.

To get the lower bound we construct the following example.
Assuming that $N$ is even, we choose a number
$k\in\{K/3,\dots,K/2\}$. Consider the configuration $X$ such that
$X(x,l)=1$ for all $x$ and $l=\{1,\dots,k\}$, while in the
$(k+1)$-th row we assume that $X(l,k+1)=0$ for odd $l$ and is
equal to $1$ otherwise. Then independently on the positions of
other particles, neither particle from the first $k$ rows can
move to the rows with numbers larger than $k$, no particles from
other rows can get into the first $k$ rows. On the other hand,
this configuration $X_{k}$ is time periodic with the period 2 and
its average velocity is equal to
$$ \v(X_{k}) = \frac{N/2}{kN + N/2} = \frac1{2k + 1}
           \le \frac1{2K/3+1} .$$
Therefore $\v(X_{k})\to0$ as $K\to\infty$.

Moreover choosing $k=1$ in the above example we get the average
velocity $\frac{N/2}{N+N/2}=1/3$ for the configuration of density
$\rho=\frac{N+N/2}{N\cdot K}=\frac3{2K}\to0$ as $K\to\infty$.
Thus for an arbitrary low density we may have the average velocity
less than $1$.

If the density $\rho>1/2$ we use the same trick as usual and pass
to the dual map and dual configuration, which permits to use the
first part of the proof to get the needed estimates.

The proof in the unbounded case is basically the same and we
leave it to the reader. \qed

There are other possible ways to model multi-row traffic flows,
for example one might consider different rules for the row
changing, the presence of high velocity particles, periodic boundary
conditions between the boundary rows, etc. However if the model
respects the law that on the next step a particle can follow its
way along the same row (assuming that the next place is not
occupied) independently on particles on neighboring rows, the
example in the above proof demonstrating the phenomenon of the
low average velocity under a low density of particles remains
generic.

\section{Passive tracer in the traffic flow}\label{sect-tracer}
In the Introduction we have mentioned the practical observation
that it is beneficial in some cases to go against the flow than
in the same direction as it goes. To study this phenomenon we
consider in this section a simple model of a passive tracer
imitating the behavior of a person moving in a hurry in the
traffic flow. As usual we assume that the tracer have its own
(forward or backward) chosen direction of motion and does not
make any impact to the flow.

Let $X(t)$ describes the 1D flow of particles and let at time $t$
the passive tracer occupies the position $x$. Then before the
next time step of the model of the flow the tracer moves in its
chosen direction to the closest (in this direction) position of a
particle of the configuration $X(t)$. For example, if the going
forward tracer occupies the position 2 and the closest particle
in this direction occupies the position 5, then the tracer moves
to the position 5. Then the next iteration of the flow occurs,
the tracer moves to its new position, etc.

To be precise for a fixed configuration $X\in\bX$ with $m(X)>1$
we introduce two maps $\tau_{X}^{+}$ and $\tau_{X}^{-}$ of the
ordered periodic lattice $\bL$ into itself defined as follows:
$$ \tau_{X}^{+}y := \min(x\in\bL: \; y<x, \; X(x)=1), $$
$$ \tau_{X}^{-}y := \max(x\in\bL: \; y>x, \; X(x)=1) $$
for any $y\in\bL$, where the order relation $y<x$ is induced by
the order on the lattice $\bL$. Then the simultaneous dynamics of
the configuration of particles (describing the flow) and the
tracer is defined by the skew product of two maps -- the map $T$
and one of the maps $\tau_{\cdot}^{\pm}$, i.e.
$$ (X,y) \to {\cal T}_{\pm}(X,y) := (TX, \tau_{X}^{\pm}y) ,$$
acting on the extended phase space $\bar \bX:=\bX\times\bL$.

Let $S(t)$ denotes the total distance covered by the tracer up to
the moment $t$ with the positive sign if the tracer moves
forward, and the negative sign otherwise. Then we define the
average (in time) velocity of the tracer $v(t)$ as $S(t)/t$.

\begin{theorem}\label{tr-th}
For an arbitrary configuration $X$ with $m(X)>1$ of the 1D model
with slow particles on the finite periodic lattice and for an
arbitrary initial configuration $X$ satisfying the regularity
assumption~(\ref{reg-ineq}) with the density $\rho \not\in
\{0,\frac12,1\}$ in the 1D unbounded case the average velocity of
the passive tracer converges to $1$ if the tracer moves forward
(along the flow)) and $\rho\le1/2$, and to $-\max(1,\frac1\rho-1)$ 
if it moves backward (against the flow).
\end{theorem}

\proof We start with the finite periodic lattice and the tracer
moving in the forward direction. According to our definition
after the first iteration of the map ${\cal T}$ the tracer will
occupy the same position as some particle. Therefore it is enough
to consider only extended configurations $(X,y)$, such that
$X(y)=1$. Since $\rho\le1/2$, then after a finite number of 
iterations the flow will consist of only free particles. Therefore 
the tracer will run down one of them and will follow it, but 
cannot outstrip. Indeed after each iteration of the flow this free 
particle occurs just one position behind the tracer.
   
We do not discuss the case $\rho>1/2$ because the average velocity 
of the tracer in this case sensitively depends on the choice of 
the initial configuration. For example in the case $n=8$ and $m=6$ 
(i.e. $\rho=3/4$) consider two initial configurations $01011111$ 
and $01111011$. In the first case the tracer after several 
iterations obtains the constant velocity $1$, while in the second 
case the velocity of the tracer is periodic with the period $3$ 
and consists of the repeating groups of $(2+2+1)$, i.e. 
the average speed is equal to $5/3$.

Consider now the case when the tracer is moving backward. Then
each time the tracer encounters a particle, on the next time step
this particle moves in the opposite direction and does not
disturb the movement of the tracer until the collision with the
next particle. Assume that the density of particles is less than
$1/2$. Then by Theorem~\ref{1D-th} after a finite number of
iterations only free particles are present in the flow, which
results in the convergence of the average velocity of the passive
tracer to $-(\frac1\rho-1)$. Indeed on the spread of length $N$
there are $m$ particles, i.e. $m$ obstacles for the tracer, which
gives the average velocity $\frac{N-m}{m}$. On the other hand, if
the density of particles is greater or equal to $1/2$ again by
Theorem~\ref{1D-th} there are no clusters of empty places in the
flow and thus after each iteration the tracer can move only by
one position, which finishes the proof for the model of the flow
with periodic boundary conditions.

The proof in the unbounded case is practically the same with the
only difference that one should use Theorem~\ref{1D-unb-th}
instead of Theorem~\ref{1D-th}. Notice that additionally to the
regularity assumption we need to assume that $\rho\not\in\{0,1\}$
to be consistent with the definition of the dynamics of the
passive tracer. \qed

Observe that the motion against the flow is efficient only in the
case of low density of particles. Certainly, this model is
oversimplified and probably its predictions are unrealistic for
the case of very high density of particles. However we believe
that while the density is not low and not very high our
description of the passive tracer is reasonable at least on the
qualitative level.

\section{Conclusion}\label{sect-conc}
This paper represents one of the first steps in the mathematical
foundation of the analysis of traffic flows and we restrict
ourselves here to the pure deterministic settings. The next step
should describe ergodic (statistical) properties of the
considered models with initial conditions chosen at random and
random versions of these models as well. This circle of questions
is especially interesting in the case of models on infinite
lattices, where the dynamics of typical configurations cannot be
obtained in the limit of infinitively large lattice sizes from
our results about regular configurations even in the
deterministic setting. Indeed, for a reasonable choice of the
class of random initial conditions their realizations do not
satisfy our regular assumption with the probability $1$.

\newpage%


\begin{thebibliography}{99}

\bibitem{Bl20} M.L. Blank, {\em Discreteness and continuity in problems
of chaotic dynamics}, Monograph, Amer. Math. Soc., 1997.

\bibitem{BF} N. Boccara, H. Fuk\'s, {\em Critical behavior of a
cellular automaton highway traffic model}, adap-org/9705003.

\bibitem{F} H. Fuk\'s, {\em Exact results for deterministic cellular
automata traffic models}, Phys. Rev. E 60 (1999), 197-202,
comp-gas/9902001.

\bibitem{KS} J. Krug, H. Spohn, {\em Universality classes for
deterministic surface growth}, Phys. Rev. A, 38 (1988),
4271-4283.

\bibitem{NH} K. Nagel, H.J. Herrmann, {\em Deterministic models for
traffic jams}, Physica A, 199 (1993), 254-269.

\bibitem{NS} K. Nagel, M. Schreckenberg, {\em A cellular
automaton model for freeway traffic}, J. Physique I, 2 (1992),
2221-2229.

\bibitem{NT} K. Nishinari, D. Takahashi, {\em Analytical properties 
of ultradiscrete Burgers equation and rule-184 cellular automaton}, 
J. Phys. A:Math. Gen., 31 (1998), 5439-5450. 

\bibitem{SN} P.M. Simon, K. Nagel, {\em Simplified cellular
automaton model for city traffic}, Phys. Rev. E, 58 (1998),
1286-1295, cond-mat/9801022.

\end{thebibliography}
\end{document}